\documentclass[proc]{revtex4}
\usepackage{graphicx}
\usepackage{fancyhdr}
\usepackage{graphics}
\usepackage{epstopdf}
\usepackage{textpos}
\usepackage{subfigure}
\usepackage{xspace}%

\newcommand{\lsp}{\PSGczDo\xspace}

\newcommand{\chiOneZero}{\ensuremath{\tilde{\chi}^{0}_{1}}\xspace}

\newcommand{\whmet}{$\PW\PH+\MET$}  
\newcommand{\njets}{\ensuremath{N_{\text{jets}}}\xspace}
\newcommand{\mht}{\ensuremath{\slash\mkern-12mu{H}_{\text{T}}}\xspace}
\newcommand{\squark}{\sQua}
\newcommand{\gluino}{\sGlu}
\newcommand{\chipmo}{\ensuremath{\tilde{\chi}_{1}^{\pm}}}
\newcommand{\chitn}{\ensuremath{\tilde{\chi}_{2}^{0}}}

\input{pdefs}

\begin{document}

\title{\centering SUSY searches in CMS}

\author{
\centering
\begin{center}
Michael Sigamani for the CMS Collaboration
\end{center}}
\affiliation{\centering University of Ghent, Belgium}
\begin{abstract}

The latest results from CMS on searches for supersymmetry are presented.
Searches involving all-hadronic final states with jets and missing transverse energy,
and in final states including one or more isolated leptons are discussed.
The results are based on 19.5 fb$^{-1}$ of LHC proton-proton collisions at $\sqrt{s}=8~$\TeV\
taken with the CMS detector.

\end{abstract}

\maketitle
\thispagestyle{fancy}

\section{Introduction}

\begin{figure}[hbt]
  \begin{center}
        \includegraphics[width=0.4\linewidth]{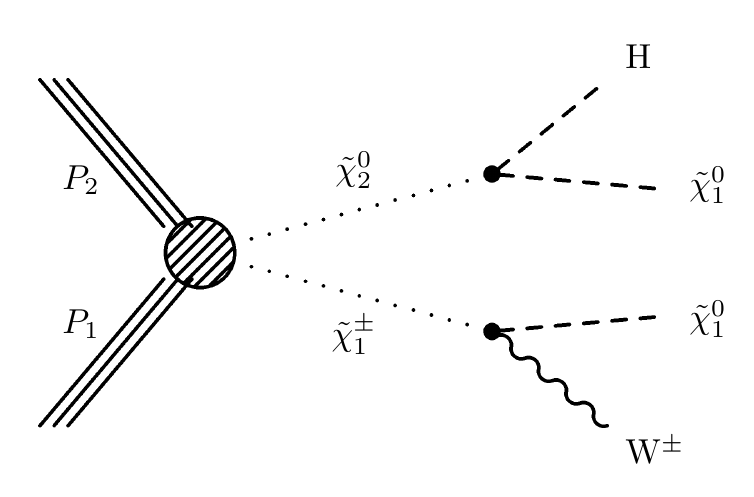}
        \includegraphics[width=0.4\linewidth]{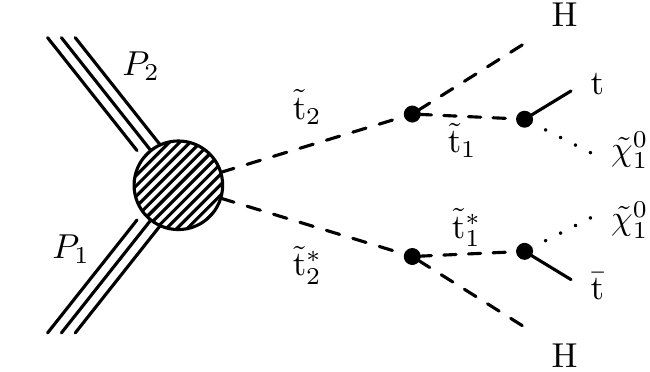}
        \includegraphics[width=0.4\linewidth]{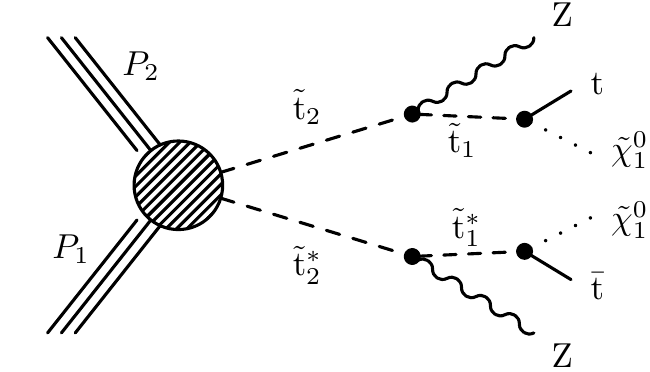}
        \includegraphics[width=0.4\linewidth]{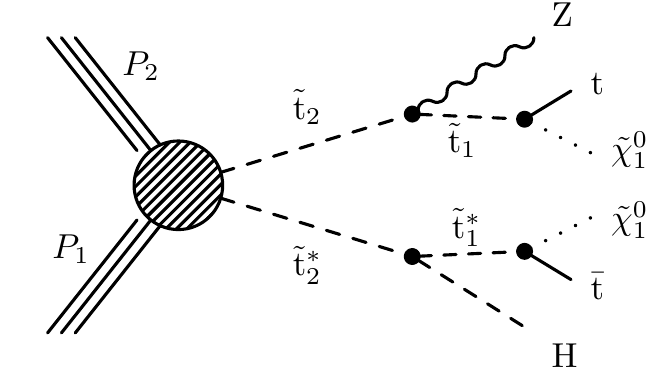}
    \caption{Diagrams for the Top (left): chargino-neutralino pair production leading to the \whmet\ final state.
             Top (right): production of the heavier top-squark ($\sttwo$) pairs followed by the decays $\sttwo\to \PH \stone$.
             Bottom (left): production of the heavier top-squark ($\sttwo$) pairs followed by the decays $\sttwo\to \cPZ \stone$.
             Bottom (left): production of the heavier top-squark ($\sttwo$) pairs followed by the decays $\sttwo\to \cPZ \PH \stone \stone$,
             with $\stone \to \cPqt \PSGczDo$. The symbol * denotes charge conjugation.
    \label{fig:SigDiagram}}
      \end{center}
\end{figure}

The standard model (SM) has been extremely successful at describing particle physics phenomena over the last half-century, 
and the recently discovered boson with a mass of 125~\GeV \cite{CMSHiggs,ATLASHiggs} could be the 
final particle required in this theory, the Higgs boson. 
However, the SM is not without its shortcomings, for instance one requires fine-tuned cancellations of large quantum corrections
in order for the Higgs boson to have a mass at the electroweak
symmetry breaking scale~\cite{SUSY1,SUSY5,SUSY6}. 
This is otherwise known as the hierarchy problem. 
Due to the magnitude of this fine-tuning, one suspects that there is some dynamical mechanism which makes this fine-tuning ``natural''.
Supersymmetry (SUSY) is a popular extension of the SM which postulates
the existence of a sparticle for every SM particle. 
These sparticles have the same quantum numbers as their SM counterparts but differ by one half-unit of spin. 
The loop corrections to the Higgs boson mass due to these sparticles 
are opposite to those of the SM particles thus providing a
natural solution to the hierarchy problem (through the cancellations of the quadratic divergences of the top-quark and top-squark loops).
Furthermore, one expects relatively light top and bottom squarks, with masses below around 1~\TeV,
if SUSY is to be the natural solution to the
hierarchy problem~\cite{Barbieri:1987fn,deCarlos1993320,Dimopoulos1995573,Papucci:2011wy}.
In addition, in $R$-parity conserving SUSY models, the lightest super-symmetric particle (LSP) is often the lightest neutralino \chiOneZero. 
The \chiOneZero offers itself as a good dark matter candidate subsequently 
explaining particular astrophysical observations~\cite{DarkMatterReview,DMGeneral}.
Since the \chiOneZero is stable this will give rise to an imbalance in transverse momentum in the detector (\MET). 
In this note, several CMS~\cite{Chatrchyan:2008aa} searches are reported for SUSY particles
using 19.5 fb$^{-1}$ of LHC proton-proton collisions at $\sqrt{s}=8~$\TeV\, 
of which the Feynman diagrams can be seen in Fig.~\ref{fig:SigDiagram}.

\section{Search for new physics with multijets and \MET}

For this study a generic search for new physics beyond the SM in
events with multiple jets and large \MET has been performed.
This final state is motivated by many extensions of the SM~\cite{LittleHiggs,UED}.
The analysis follows previous
inclusive searches~\cite{RA2,RA2_2011} which required at least three jets in
the final state. These searches were sensitive to the simple decay chains of
squarks and gluinos. In order to give the analysis more sensitivity to a variety of final
state topologies the events are sub-divided into
three exclusive jet multiplicity regions: \njets=[3--5], [6--7], and [$\geq$8].
This extension of the search to higher jet multiplicities
is motivated by natural SUSY models whereby the gluino decays often into top quarks.
While other analyses exploit the presence of b jets in signal events to discriminate against
background \cite{RA1_2012, RA2b_2012}, this analysis follows a complementary strategy by requiring a large number of jets,
and thus is keeping the signal efficiency for fully hadronic final states as high as possible.
The events are further categorized according to the total visible hadronic activity \HT and the
momentum imbalance \mht, defined in the direction transverse to the beam.

\begin{figure}
  \centering
    \subfigure[Squark pair production (T2qq)]{
     \includegraphics[width=0.45\textwidth] {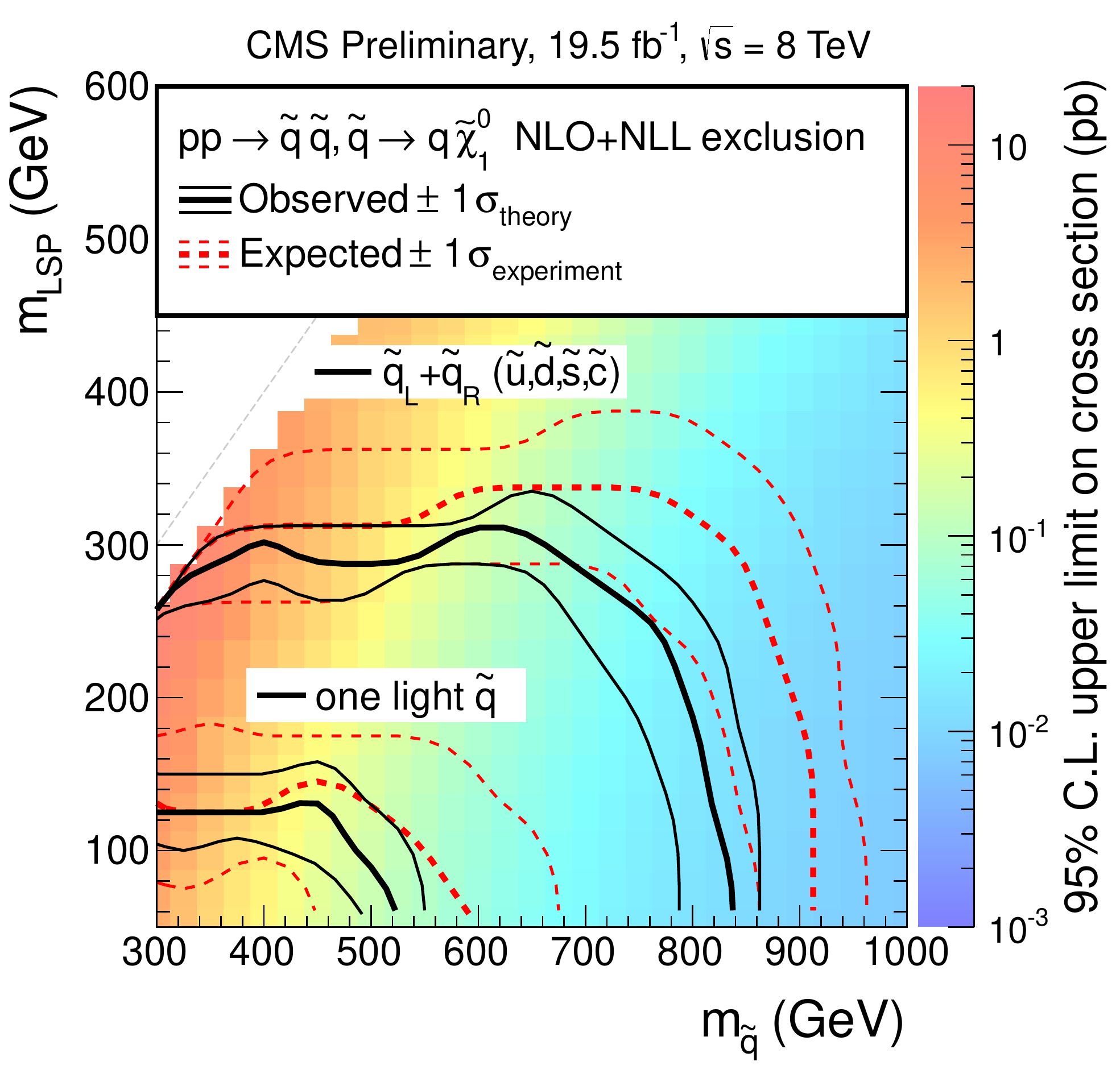}
     \label{fig:limitsT2qq}
    }
    \subfigure[Gluino pair production (T1qqqq)]{
     \includegraphics[width=0.45\textwidth] {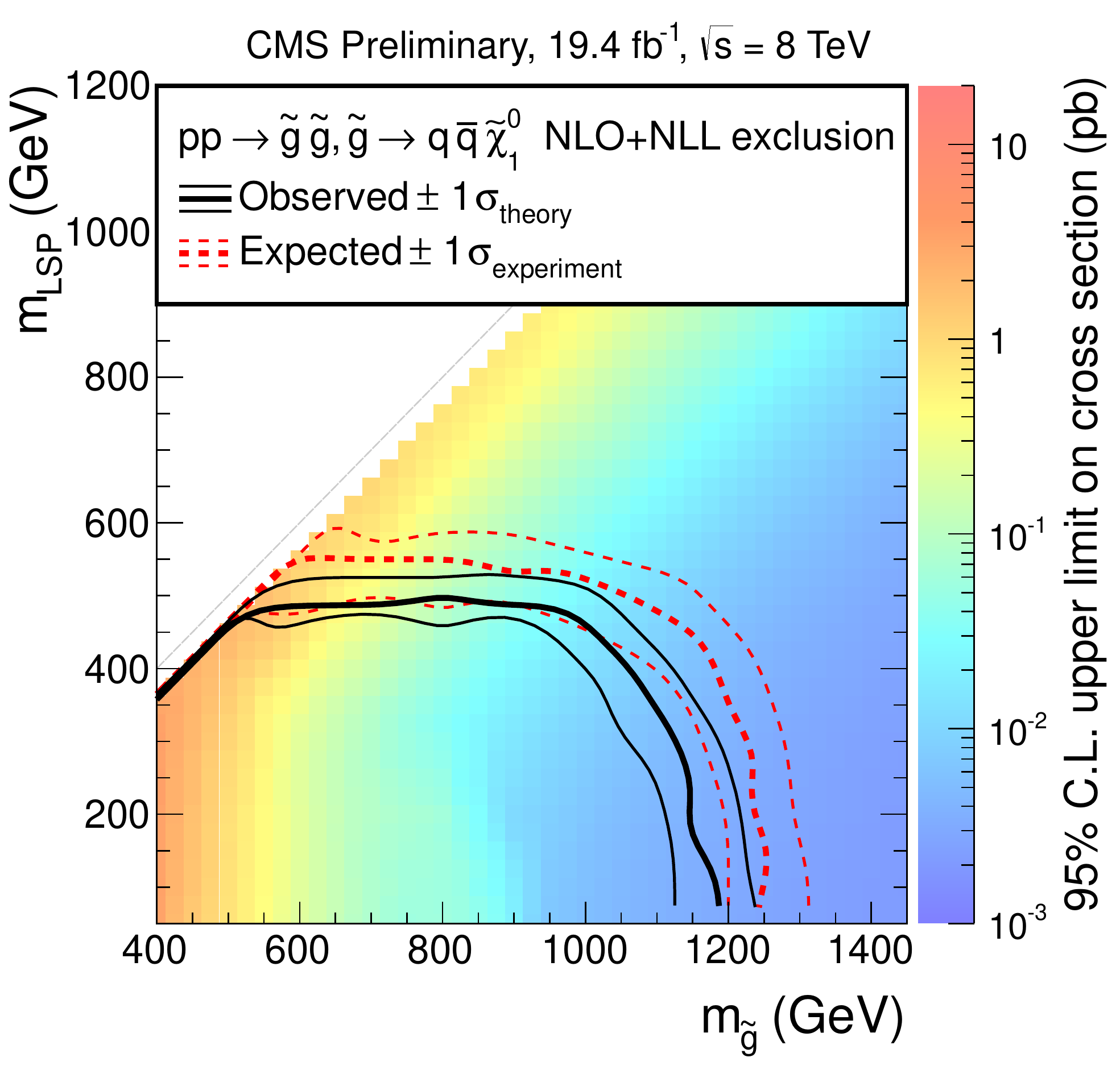}
     \label{fig:limitsT1qqqq}
    }
  \caption{The observed and expected 95\% C.L. upper limits on the (a) $\squark\squark$ and (b) $\gluino\gluino$
    production cross-sections in the $m_{\squark}$-$m_{\chiz}$
and $m_{\gluino}$-$m_{\chiz}$ planes obtained with the simplified models. For the $\squark\squark$ production the upper set of curves corresponds to the
scenario when the first two generations of squarks are degenerate and light, while the lower set corresponds to only one light accessible squark.}
  \label{fig:limitsT1qqqqT1tttt}
\end{figure}

The major SM background contributions in this search arise from
an irreducible background of \Z{}+jets events, with the \Z boson decaying to a pair of neutrinos, denoted as
$\znunubr$+jets; $\W$+jets and $\ttbar$ events, with a $\W$ boson decaying directly or via a
$\tau$ to an e or $\mu$ that is not reconstructed, isolated, or out of acceptance; or to a $\tau$ that decays hadronically.
In all these cases, one or more neutrinos provide a source of genuine \MET
The third background category is QCD multijet events with
large \MET from leptonic decays of heavy-flavour hadrons
inside the jets, jet energy mismeasurement, or instrumental noise and
non-functioning detector components. All these backgrounds are determined
from the data, relying on simulation as little as possible.
The full details of the event selection with analysis procedure can be found in Ref.~\cite{SUS13012}.
After all selections have been made one finds the observed data are consistent with SM expectations
and subsequently limits are placed on the production of $\squark\squark$ pairs
with $\squark \rightarrow \mathrm{q} + {\chiz}$, and $\gluino\gluino$ pair with
$\gluino \rightarrow \qqbar + {\chiz}$ in the $m_{\squark}$-$m_{\chiz}$ and $m_{\gluino}$-$m_{\chiz}$ planes (see Fig.~\ref{fig:limitsT1qqqqT1tttt}).

\section{Search for production of charginos and neutralinos with a Higgs boson}
\begin{figure}[!h]
\begin{center}
\includegraphics[width=0.7\textwidth]{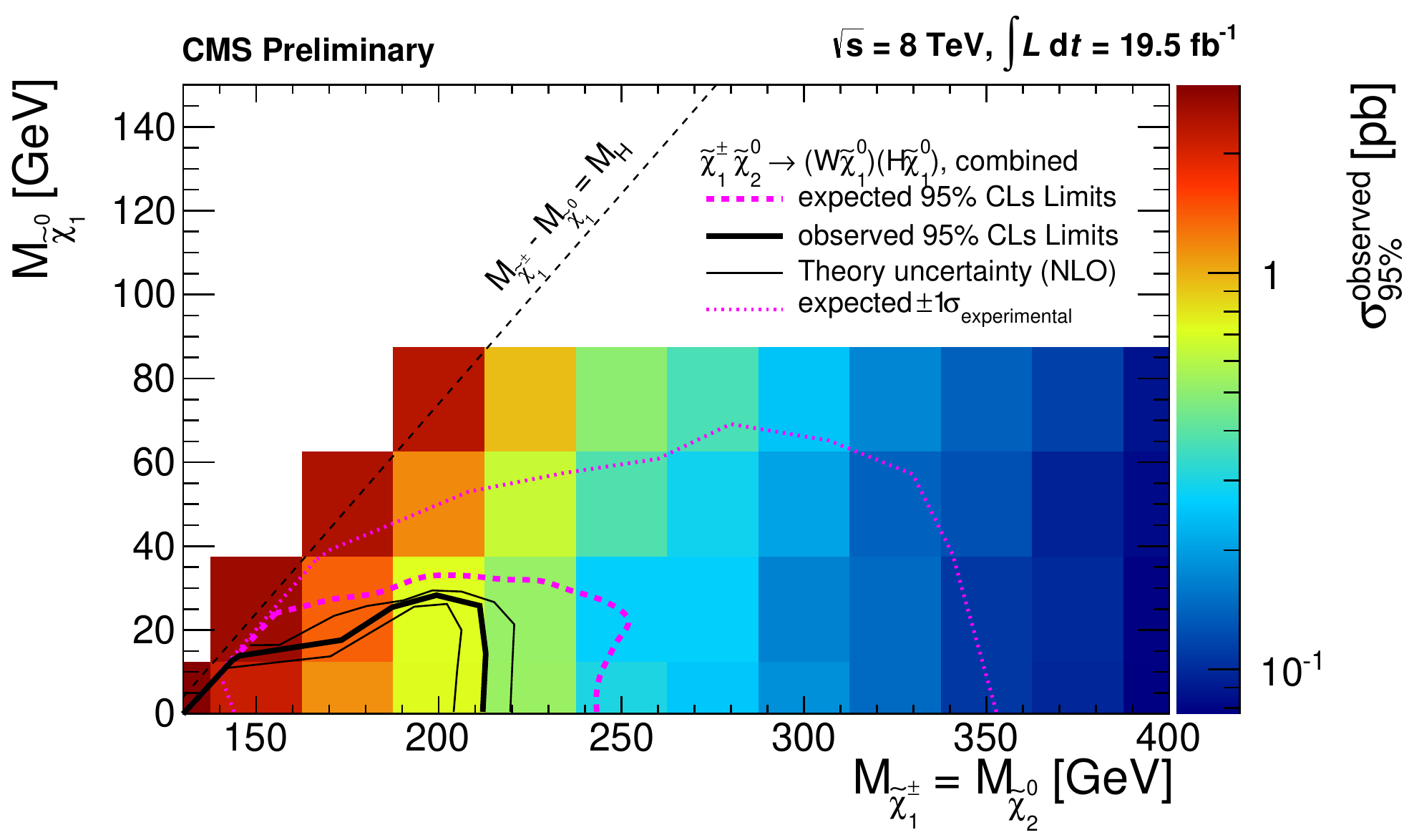}
\caption{ The interpretation of the combined results from the three search channels.
The upper limit on the \chipmo\chitn\ production cross section is indicated in the color scale.
The expected and observed regions for which the signal model is excluded reach
from the origin to the solid red and solid black curve, respectively.
The dashed red lines show the $\pm 1\sigma$ variations on the expected limit due to experimental uncertainties
and the thin black lines indicate the uncertainty due to the cross section calculation.
\label{fig:interpretations2d}
}
\end{center}
\end{figure}

The recent observation of a Higgs boson offers the
possibility to perform new physics searches by exploiting the measured properties
of this particle. In particular, in large regions of SUSY parameter space the heavy neutralinos
are expected to decay predominantly to a Higgs boson.
The electroweak SUSY process with the largest cross section is chargino-neutralino pair production.
This process has been probed in previous CMS searches~\cite{Chatrchyan:2012pka,CMS-PAS-SUS-13-006}, which required that
the chargino decays to a \PW\ boson and the lightest SUSY particle (LSP), assumed to be the
lightest neutralino $\lsp$, and that the neutralino decays to a \Z boson and the LSP.
The search is based on chargino-neutralino pair production where the neutralino decays instead
to a Higgs boson and the LSP, $\chipmo\chitn\to(\PW^{\pm}\lsp) (\PH \lsp)$, as shown in Fig.~\ref{fig:SigDiagram} (top left).
The decays $\chipmo\to\PW^{\pm}\lsp$ and $\chitn\to \PH\lsp$ are expected to dominate if the
$\chipmo$ and $\chitn$ particles are wino-like, the $\lsp$ is bino-like, and the difference
between their masses is larger than $M_{\PH}=126$\GeV.
Here the $\PH$ particle is the lightest SUSY Higgs boson, which is expected to be SM-like
if the other SUSY Higgs bosons are significantly heavier than $M_{\Z}$~\cite{SUSYprimer}.
Three exclusive final states are considered in this search, where the \PW\ is required
to decay leptonically and the $\lsp$ is assumed to be stable and escape detection, leading to large \MET.
A search performed in the single lepton final state provides
sensitivity to events in which the Higgs boson decays as $\PH\to\cPqb\bar{\cPqb}$. A search in the same-sign dilepton
final state targets events with the decay $\PH\to\PW^+\PW^-$ in which one of the \PW\ bosons
decays leptonically and the other hadronically.
The full details of the event selection with analysis procedure can be found in Ref.~\cite{CMS:2013afa}.
After all selections have been made one finds the observed data are consistent with SM expectations
and subsequently limits are presented in the plane of the mass of the \chiOneZero,
and the common mass of the $\chipmo$ and $\chitn$ particles (see Fig.~\ref{fig:interpretations2d}).

\section{Search for top-squark pair production with Higgs and Z bosons} 

Searches for direct top-squark production at the LHC have focused mainly
on the simplest scenario, in which only the lighter top-squark
mass eigenstate, $\stone$, is accessible at current collision energies. 
In these searches, the top-squark decay modes
considered are those to a top quark and a neutralino,
$\stone\to \cPqt\PSGczDo \to \cPqb \PW \PSGczDo$, or
to a bottom quark and a chargino, $\stone\to \cPqb \PSGcpDo \to \cPqb \PW \PSGczDo$.
These two decay modes are expected to have large branching fractions
if kinematically allowed.
The lightest neutralino, $\PSGczDo$, is the lightest
SUSY particle (LSP) in the R-parity conserving models
considered; the experimental signature of such a particle is
missing transverse energy ($\MET$).

The sensitivity of searches for direct top-squark pair production is, however, significantly reduced
in the $\stone\to \cPqt\PSGczDo$ decay mode for the region of SUSY parameter space in which
 $m_{\stone} - m_{\PSGczDo} \simeq m_{\cPqt}$.  
In this region, the momentum of the daughter neutralino in the rest frame of the decaying
$\stone$ is small, and it is exactly zero in the limit $ m_{\stone} - m_{\PSGczDo} = m_{\cPqt}$. As a result, the $\MET$
from the vector sum of the transverse momenta of the two neutralinos is typically also small in the laboratory frame.
It then becomes difficult to distinguish kinematically between
$\stone$ pair production and the dominant background,
which arises from $\ttbar$ production.
This region of phase space can be explored using events with topologies that are distinct from the $\ttbar$ background.
An example is gluino pair production where each gluino decays to a top squark and a top quark, giving rise to a
signature with four top quarks in the final state~\cite{Chatrchyan:2013iqa,Chatrchyan:2013wxa}.

This search targets the region of phase space where
$m_{\stone} - m_{\PSGczDo} \simeq m_{\cPqt}$ by focusing on signatures
of $\ttbar\PH\PH$, $\ttbar\PH\cPZ$, and
$\ttbar\cPZ\cPZ$ with $\MET$. These final states can
arise from the pair production of the heavier top-squark mass eigenstate
$\sttwo$. There are two non-degenerate top-squark mass eigenstates
($\sttwo$ and $\stone$) due to the mixing of the SUSY partners $\stL$ and $\stR$ of the
right- and left-handed top quarks. The $\sttwo$
decays to $\stone$ and an $\PH$ or $\cPZ$ boson,
and the $\stone$ is subsequently assumed to decay to $\cPqt\PSGczDo$, as
shown in the bottom row of Fig.~\ref{fig:SigDiagram}. 
The final states pursued in this search can arise in other scenarios, such as $\stone\to
\cPqt\PSGczDt$, with $\PSGczDt \to \PH
\PSGczDo$ or $\PSGczDt \to \cPZ
\PSGczDo$. The search is also sensitive to a range of models in
which the LSP is a gravitino.
The relative branching fractions for modes with the $\PH$ and $\cPZ$ bosons are
model dependent, so it is useful to search for both decay modes simultaneously.
In the signal model considered, $\sttwo$ is assumed always to decay to
$\stone$ in association with an $\PH$ or $\cPZ$ boson,
such that the sum of the two branching fractions is
$\mathcal{B}(\sttwo\to \PH \stone) + \mathcal{B}(\sttwo\to \cPZ
\stone) = 100$\%. Other possible decay modes are $\sttwo \to
\cPqt\PSGczDo$ and $\sttwo \to \cPqb \PSGcpDo$. 

\begin{figure}[htbp]
\centering
\includegraphics[width=0.49\textwidth]{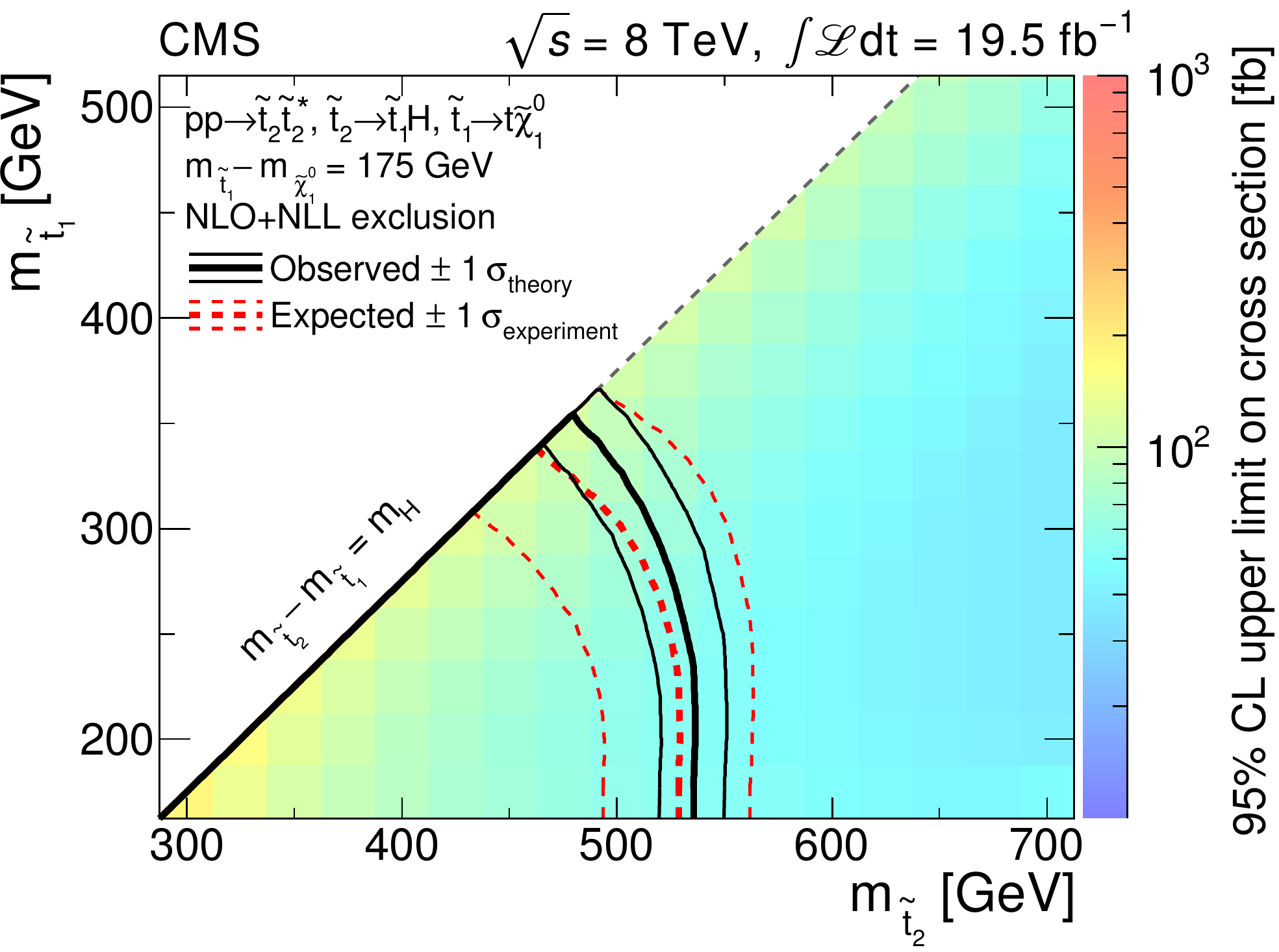}
\includegraphics[width=0.49\textwidth]{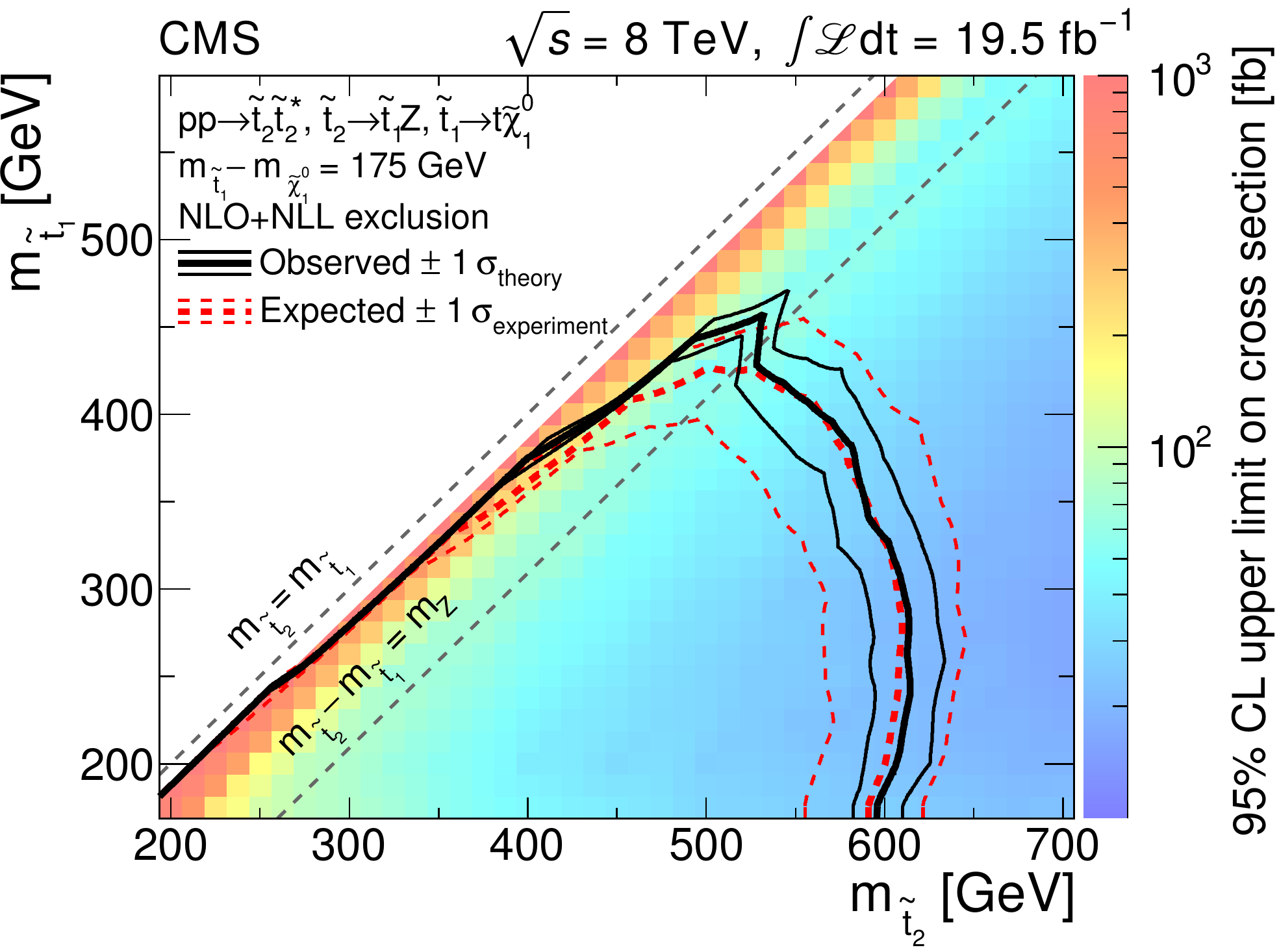}
\caption{
  Interpretation of the results in SUSY simplified model
  parameter space, $m_{\stone}$ \vs $m_{\sttwo}$, with the
  neutralino mass constrained by the relation
  $m_{\stone} - m_{\PSGczDo}  = 175\GeV$.
  The shaded maps show the upper
  limit (95\% CL) on the cross section times branching fraction
  at each point in the $m_{\stone}$ \vs~$m_{\sttwo}$ plane for the
  process $\Pp\Pp \to \sttwo\sttwo^{*}$,
  with $\sttwo\to \PH \stone$, $\stone\to
  \cPqt\PSGczDo$ and $\PSQtDt\to \cPZ\PSQtDo$,
  $\stone\to \cPqt\PSGczDo$. In these plots,
  the results from all channels are combined. The excluded region
  in the $m_{\stone}$ \vs $m_{\sttwo}$ parameter space is obtained
  by comparing the cross section times branching fraction upper
  limit at each model point with the corresponding NLO+NLL
  cross section for the process, assuming that (a)~$\mathcal{B}(\sttwo\to \PH \stone)=100\%$ or (b)~that $\mathcal{B}(\sttwo\to \cPZ \stone)=100\%$.
  \label{fig:interp_HH_and_ZZ}}
\end{figure}

The four main search channels contain either exactly
one-lepton, two leptons with opposite-sign (OS) charge and no other
leptons, two leptons with same-sign (SS) charge and no other leptons,
or at least three leptons (3 $\ell$).
The channels with one-lepton or two OS leptons require at least three
$\cPqb$ jets, while the channels with two SS leptons or 3 $\ell$ require
at least one $\cPqb$ jet.
The major SM background contribution in this search arise from $\ttbar$ pair production,
which has two $\cPqb$ quarks and either one-lepton or two OS leptons from
the $\ttbar\to \ell \nu \qqbar \bbbar$ or
$\ttbar\to \ell \nu \ell \nu \bbbar$ decay modes, where
$\cPq$ denotes a quark jet. The
sensitivity to the signal arises both from events with additional
$\cPqb$ quarks in the final state (mainly from
$\PH\to\bbbar$), and from events with additional leptons from $\PH$ or $\cPZ$ boson decays.
The full details of the event selection with analysis procedure can be found in Ref.~\cite{Chachatryan:2014doa}. 
The observed data yields are consistent with SM expectations
and subsequently limits are placed on top quark pair production (see Fig.~\ref{fig:interp_HH_and_ZZ}).

\section{Phenomenological MSSM Interpretation of 7 and 8 TeV results}

Searches for SUSY have typically been interpreted within constrained models with just a few parameters.
A commonly used model is the Constrained Minimal Supersymmetric Standard Model (CMSSM),
which is characterized by four parameters and a sign~\cite{Chamseddine:1982jx}:
a universal scalar mass $m_0$, a gaugino mass $m_{1/2}$ and a trilinear coupling $A_0$
defined at the GUT scale, $M_{\rm GUT}\sim 10^{16}$~GeV,
together with $\tan\beta$ and sign$(\mu)$.
The simplifying assumption of universality at the GUT scale has served a useful purpose:
for many years it has provided a framework for gauging progress in SUSY searches.
However, many mass patterns and signatures that are possible in the MSSM
cannot be realized in the CMSSM. Therefore, interpreting the experimental results only in the $(m_0,m_{1/2})$ plane
carries the risk of imposing overly strong constraints on SUSY that are not warranted by observations.
The full set of mass patterns and signatures possible in the MSSM are also not necessarily accounted for in the
Simplified Model Spectra (SMS)~\cite{Alwall:2008ag,Alves:2011wf,Chatrchyan:2013sza} approach.
Although interpretation using SMS topologies is very useful since it allows us to systematically see the experimental impact on well-defined,
isolated topologies, it is crucial to complement this by interpretation within a generic model,
such as the phenomenological MSSM (pMSSM)~\cite{Djouadi:1998di}, that intrinsically covers a wide diversity of topologies.
Of course, it is also possible that the SMS approach eliminates a pMSSM parameter point that is not eliminated in the present
version of the analysis presented here if the latter set of analyses does not include some important channel/configuration considered by the SMS analyses.
The SMS approach and the pMSSM approach become equivalent only when: a)  the set of analyses in the pMSSM approach is the same as the set of analyses
considered in the simplified models; and b) the simplified models cover all the final states to which these analyses can be sensitive for each pMSSM point.
In order to account fully for all the mass patterns and decay modes that can occur in the MSSM,
it is necessary to pursue a less model-dependent approach.
Subsequently, one can pursue the pMSSM, a 19-dimensional realization of the MSSM,
which captures most of the phenomenological features of the R-parity conserving MSSM.
In the pMSSM, all MSSM parameters are specified at the electroweak scale and allowed to vary freely subject to the requirement that the model be
consistent with electroweak symmetry breaking and other such basic constraints.

To assess what the data tells us about SUSY in the context of the pMSSM
a representative subset of the results obtained by CMS corresponding to integrated luminosities of 5.0~fb$^{-1}$ at 7~TeV and  19.5~fb$^{-1}$ at 8~TeV is used.
The current study extends our analysis \cite{sus12030} of the 7~TeV data.
The study uses a subspace of the pMSSM where the chargino lifetime $c\tau(\tilde{\chi}^\pm_1)$ is less than 10 mm to look at the class of final states with prompt decays.
In order to combine both data sets the same model points within the pMSSM parameter space are used, chosen randomly from a scan of points
consistent with basic constraints and treat the 7~TeV and 8~TeV data in an entirely parallel fashion.
The approach employed is an extension of the work from Ref.~\cite{Sekmen:2011cz}.
Prior to this work, the parameter space of the pMSSM and the various LHC constraints on it were studied in detail in
Refs.~\cite{Berger:2008cq,AbdusSalam:2009qd,Conley:2010du}.

\subsection{Posterior densities for parameters, masses and relevant observables}

\begin{figure}[htbp]
  \centering
  \includegraphics[width=0.32\linewidth]{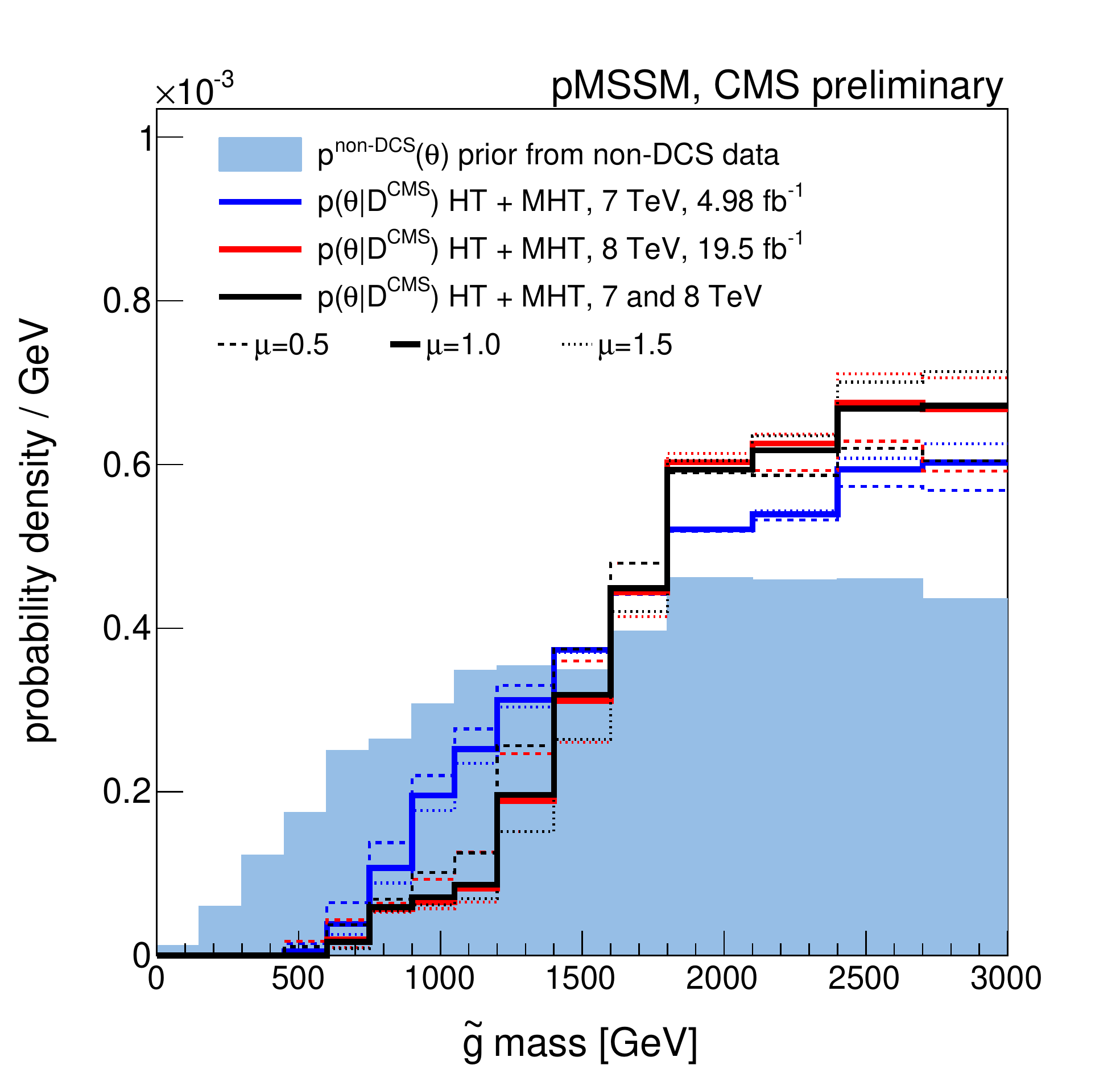}
  \includegraphics[width=0.32\linewidth]{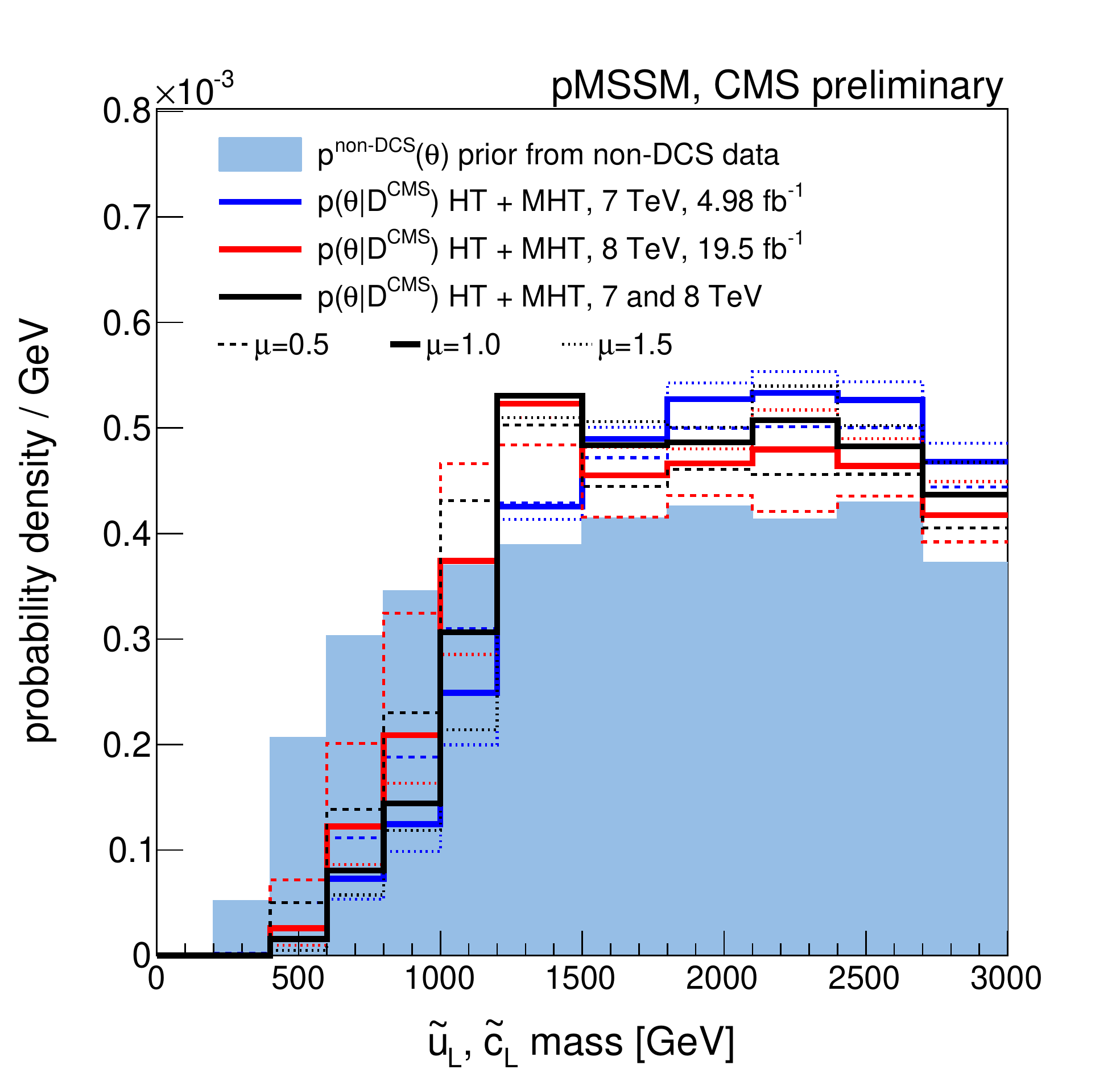}
  \includegraphics[width=0.32\linewidth]{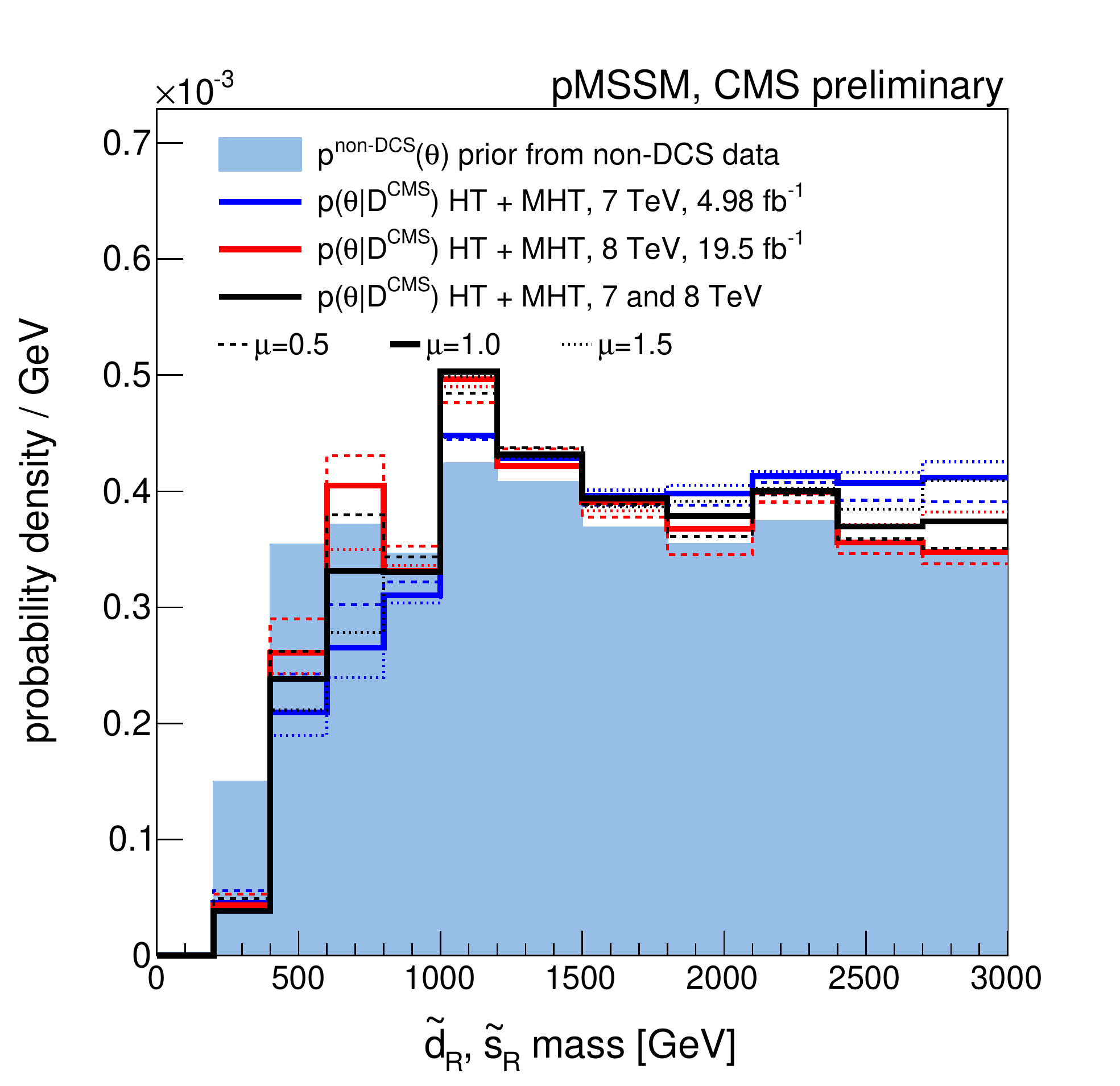}
  \includegraphics[width=0.32\linewidth]{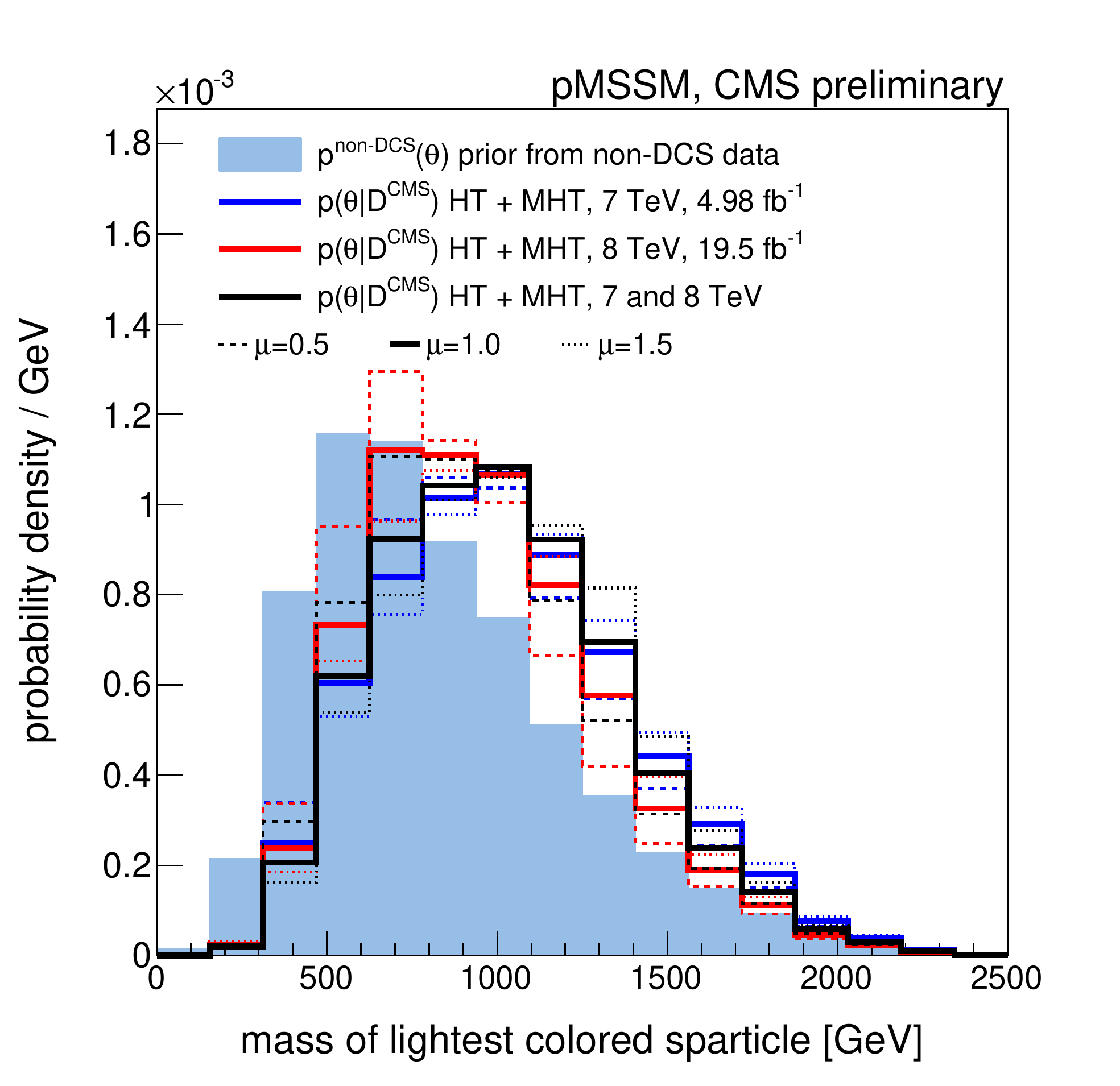}
  \includegraphics[width=0.32\linewidth]{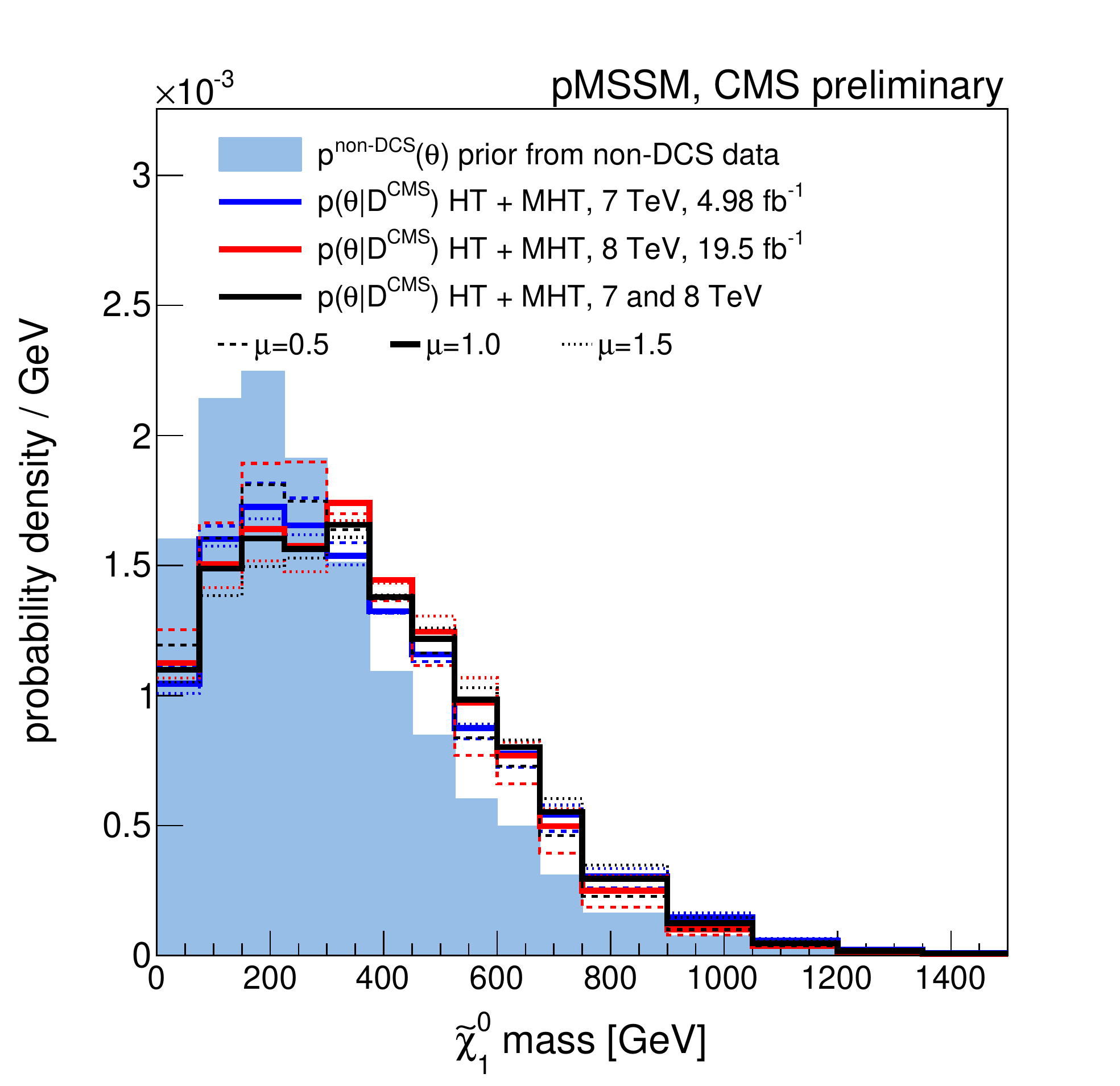}
  \caption{Marginalized distributions of selected sparticle masses. Filled histograms show prior
distributions, line histograms show posterior distributions including the data collected at 7 TeV
(blue), 8 TeV (red), and 7 and 8 TeV (black) by the CMS HT + MHT searches.}
  \label{fig:HT_MHT_1D}
\end{figure}

Fig.~\ref{fig:HT_MHT_1D} compares prior distributions to posterior distributions including the data from the 7 and 8~TeV HT + MHT searches performed by CMS~\cite{SUS12011,SUS13012}.
Blue line histograms represent marginalized CMS posterior distributions. 
The red and black line histograms show the similar CMS posterior distributions for respectively the 8~TeV and 7+8~TeV HT+MHT data, where the 7+8~TeV combined posterior probability
for each point is obtained by taking a product of the 7~TeV and 8~TeV likelihoods for this search.
Solid lines show posterior distributions assuming the central values for the signal cross section,
while dashed and dotted lines show posterior distributions assuming respectivel 0.5 and 1.5 times the central values for the signal cross section.
The difference between the solid lines and the dotted and dashed lines can be considered as a systematic uncertainty.

It appears that the HT+MHT data strongly disfavor pMSSM scenarios with $\gluino$ masses below 1200 GeV.
Also scenarios with $\tilde{u}_L,\tilde{c}_L$ masses below 1000 GeV are disfavored.
However, the impact on masses of other first and second generation squarks, such as the $\tilde{d}_{R},\tilde{s}_{R}$ masses, is weaker.
Regarding third generation squarks, there is a slight impact on the mass of the lightest sbottom, disfavoring the lowest masses,
while there is no noticeable impact on the mass of the lightest stop.
Generally, the most probable mass of the lightest colored sparticle is increased by about 500~GeV.
Also the distribution of the $\chiOneZero$ mass is shifted to higher values.
This latter effect is mainly a phase space effect, a consequence of our requirement of $\chiOneZero$ to be the LSP.
The full study can be found in Ref.~\cite{CMS:2014mia}.

\section{Summary} 

CMS has carried out several searches for SUSY particles with an 
integrated luminosity of 19.5 fb$^{-1}$. No excess in data with respect to the SM expectation has been observed so far.
However, searches in large regions of the parameter space for natural SUSY are still
in progress. These searches are challenging due to similarity with the $\ttbar$ final state for low stop masses, and due to the low cross sections for higher
stop mass values. Subsequently, new results obtained at $\sqrt{s}=13~$\TeV\ in 2015 will be looked upon with eager anticipation.

\bigskip 

\bibliography{}

\end{document}